\documentclass[conference]{IEEEtran}
\IEEEoverridecommandlockouts
\usepackage{cite}
\usepackage{amsmath,amssymb,amsfonts}
\usepackage{algorithmic}
\usepackage{amsthm}
\usepackage{graphicx}
\usepackage{textcomp}
\usepackage{xcolor}
\usepackage{verbatim}
\usepackage{physics} % braket
\usepackage{commath}
\usepackage{mathtools}
\usepackage{enumitem}
\usepackage{multirow}
\usepackage{booktabs}
\usepackage{hyperref}
\usepackage[margin=0.72in]{geometry}
\usepackage[export]{adjustbox}
\DeclarePairedDelimiter{\ceil}{\lceil}{\rceil}
\theoremstyle{definition}
\newtheorem{definition}{Definition}
\newtheorem{example}{Example}
\newtheorem{theorem}{Theorem}
\newtheorem{proposition}{Proposition}
\newcommand\scalemath[2]{\scalebox{#1}{\mbox{\ensuremath{\displaystyle #2}}}}
\newcommand{\overbar}[1]{\mkern 1.5mu\overline{\mkern-1.5mu#1\mkern-1.5mu}\mkern 1.5mu}
\def\BibTeX{{\rm B\kern-.05em{\sc i\kern-.025em b}\kern-.08em
    T\kern-.1667em\lower.7ex\hbox{E}\kern-.125emX}}
\begin{document}

%\title{BDD-based Quantum Simulation\\}
\title{Bit-Slicing the Hilbert Space: Scaling Up Accurate Quantum Circuit Simulation to a New Level
%\thanks{This work was supported in part by the Ministry of Science and Technology of Taiwan under grant MOST 108-2218-E-002-073.}
}

%\iffalse
\author{\IEEEauthorblockN{Yuan-Hung Tsai$^\dag$, Jie-Hong R. Jiang$^{\dag\ddag}$, and Chiao-Shan Jhang$^{\ddag}$}
\IEEEauthorblockA{\textit{$^\dag$Graduate Institute of Electronics Engineering, $^\ddag$Department of Electrical Engineering} \\
\textit{National Taiwan University}, Taipei, Taiwan
}
}
%\fi

\maketitle

\begin{abstract}
Quantum computing is greatly advanced in recent years and is expected to transform the computation paradigm in the near future.
Quantum circuit simulation plays a key role in the toolchain for the development of quantum hardware and software systems.
However, due to the enormous Hilbert space of quantum states, simulating quantum circuits with classical computers is extremely challenging despite notable efforts have been made.
In this paper, we enhance quantum circuit simulation in two dimensions: accuracy and scalability.
The former is achieved by using an algebraic representation of complex numbers; the latter is achieved by bit-slicing the number representation and replacing matrix-vector multiplication with symbolic Boolean function manipulation.
Experimental results demonstrate that our method can be superior to the state-of-the-art for various quantum circuits and can simulate certain benchmark families with up to tens of thousands of qubits.
%Quantum computation is an emerging research field, which provides considerable speed-ups when conducting important applications.
%However, since the implementation of real quantum computers is still immature, simulation of quantum computation on classical computers is highly in demand in this research field.
%The main bottleneck of quantum simulation is its exponential complexity, which results from vector and matrix representations for quantum states and operations.
%As the sizes of vectors/matrices grow exponentially with respect to the number of qubits, traditional methods with array-based representations can only deal with rather small quantum circuits.
%Some existing solutions tackle this problem with decision diagrams, which are rather compact in size compared to array-based approaches, and obtain a huge improvement.
%In this paper, we come up with a new bitwise representation for quantum simulation based on binary decision diagrams.
%We derive Boolean formulas for each quantum operations to avoid the vector-matrix multiplication. By reordering variables during the simulation process, the binary decision diagrams are possible to retain in acceptable sizes.
%Experimental results show that the proposed method outperforms the state-of-the-art in terms of running time, memory usage, and accuracy.
\end{abstract}

%\begin{IEEEkeywords}
%quantum computation, quantum simulation, quantum simulator, binary decision diagram
%\end{IEEEkeywords}

\section{Introduction}
Recent progress in building quantum computers has set the milestone of demonstrating quantum supremacy \cite{supremacy19}.
Quantum computation is expected to provide computing power beyond the reach of classical computers and transform the information technology in the near future.
Quantum hardware and software systems, e.g., \cite{IBMQ,Qiskit}, are under active development.
Quantum system design requires a comprehensive software toolchain, where quantum circuit simulation is one of the key components.

Simulating quantum circuits on a classical computer is indispensable to understand system behavior and verify design correctness especially before universal quantum computers are ready.
However the simulation is challenging because quantum states have to be described in the complex vector space and the space is exponential in the number of quantum bits (qubits).
Although there are special classes of quantum circuits, such as the stablizer circuits, that allow efficient simulation by classical computers, see, e.g., \cite{Aaronson_2004}, simulating general quantum circuits can be extremely difficult.
In fact, it is this difficulty that triggered Richard Feynman envisaging quantum simulators/computers in his seminal lecture on 'Simulating Physics with Computers' in 1981.

Despite the aforementioned challenges, various simulation algorithms have been proposed and tools are available to date.
Depending on the underlying data structure, existing methods can be mainly classified into array-based, e.g., \cite{Quipper,liqui,QX,breaking,ProjectQ} , or decision-diagram (DD) based, e.g., \cite{QCS,MTBDD,zulehner2019advanced,QMDD2}.
The former is rather limited without exploiting supercomputing facilities and hardly scalable to 50 qubits even with supercomputing.
On the other hand, although decision diagrams are well-known for their typical memory explosion problems, the latter when engineered properly can be superior to the former \cite{zulehner2019advanced}.
The simulation method proposed in this work is
DD-based.
%using off-the-shelf binary decision diagram (BDD) packages.
While prior DD-based methods \cite{QCS,QMDD1,QMDD2} require specializing multi-terminal or multi-valued DDs for general quantum circuit simulation, ours relies on standard binary decision diagrams (BDDs) and takes an off-the-shelf BDD package for computation.

The state-of-the-art methods \cite{zulehner2019advanced,QMDD2} are based on the Quantum Multiple-valued Decision Diagrams (QMDDs) \cite{QMDD1}.
The data structure consists of decision nodes with multi-valued branching for matrix representation and edges weighted with complex numbers for unitary operator and state vector representation and manipulation.
In contrast, we simply rely on BDD to represent quantum states and support matrix and vector multiplication.
Moreover, unlike prior work \cite{zulehner2019advanced,QMDD2} with precision loss representing complex numbers, our method employs the algebraic representation \cite{accuracy} for accurate complex number representation under the considered set of unitary operators (see Table~\ref{tb:gate}) general enough to achieve universal quantum computation.
Note that although the QMDD-based methods can potentially benefit from algebraic representation, it cannot be done as easy as ours due to the complications of unique representation and division by normalization factors \cite{accuracy}.
To the best of our knowledge, our method is the first work that utilizes the accurate representation for quantum circuit simulation.

In addition to the accuracy enhancement, to extend the capacity of quantum circuit simulation, we devise 1) a bit-slicing technique that represents a state vector bit by bit each corresponds to a BDD, and 2) an implicit method that replaces matrix-vector multiplication with a set of precharacterized Boolean formulas of the unitary operators for BDD manipulation.
Experimental results demonstrate the accuracy and scalability advantages of the proposed method compared to the state-of-the-art over a number of different benchmarks.
Notably for certain benchmark families, our method can simulate circuits up to tens of thousands of qubits beyond the capacity of other existing simulators.
While encouraged by the strengths of our approach, we also identify some weaknesses for future improvements.

%BDD as underlying data structure
%describe the advantage of using BDD
%talk about it is easy to conduct boolean operations on BDDs
%talk about derive boolean formulas
%universal:Accuracy and Compactness: any quantum operation can be realized exactly or up to an arbitrarily small error

The rest of the paper is organized as follows.
In Section~\ref{sec:prelim}, preliminaries are given.
The main simulation algorithm is then presented in Section~\ref{sec:alg}.
Section~\ref{sec:exp} shows the experimental results and evaluates different simulation methods.
Finally in Section~\ref{sec:concl} we concludes this paper and outline some directions for future work.

\section{Preliminaries} \label{sec:prelim}
For convention in the sequel, variables are denoted with lower-case letters, e.g. $x$, while Boolean functions and their BDD representations are denoted with upper-case letters, e.g., $F$.
We denote Boolean connectives negation by overline or $\neg$, conjunction by $\wedge$, disjunction by $\vee$, and exclusive-or by ${\oplus}$.
We sometimes omit $\wedge$ in a Boolean formula.

A \emph{literal} is a Boolean variable, e.g., $x$, in the  positive phase, or its negation, e.g., $\neg x$, in the negative phase.
Let $\mathit{var}(\ell)$ denote the underlying variable of literal $\ell$; let $\mathit{phase}(\ell)$ denote the phase of $\ell$ for $\mathit{phase}(\ell) = 1$ if $\ell = \var(\ell)$ and $\mathit{phase}(\ell) = 0$ otherwise.
A \emph{cube} is a conjunction of literals, which is treated as a set of literals.

The \emph{cofactor} of a Boolean function (BDD) $F$ with respect to a literal $\ell$ is denoted by $F|_\ell$, which corresponds to the new Boolean function same as $F$ expect for variable $\mathit{var}(\ell)$ in $F$ being substituted with $\mathit{phase}(\ell)$.
The notion of cofactor is straightforwardly generalized to a cube $q$ so that $F$ is cofactored with respect to the literals in $q$.

%integer vector, complex value vector, bit vector?
%Probability?
%vector, matrix?

\subsection{Quantum Circuit Basics} \label{sec:qcb}
Quantum computation through quantum circuit execution takes three actions: 1) initial state preparation, 2) state evolution via quantum circuit operation, 3) qubit measurement.
A quantum circuit simulator has to implement algorithms to perform these actions as detailed in the following.

\subsubsection{Initial State Preparation}
Unlike a classical bit takes value either 0 (in other words, in state 0, denoted $\ket{0}$) or 1 (in state 1, denoted $\ket{1}$), a qubit in state $\ket{\psi}$ can be in a superposition state of both $\ket{0}$ and $\ket{1}$ described by
\[
\ket{\psi} = \alpha\cdot\ket{0} + \beta\cdot\ket{1}
\]
where $\alpha, \beta \in \mathbb{C}$ are \emph{probability amplitudes} satisfying the normalization constraint ${\abs{\alpha}}^{2}+{\abs{\beta}}^{2}=1$.
For an $n$-qubit quantum system, the qubits can be entangled and an $n$-qubit state $\ket{\psi}$ can be described by
\begin{equation}
\ket{\psi}=\displaystyle\sum_{i\in{\{0, 1\}}^{\it n}} \alpha_{i}\cdot\ket{i},
\label{eq1}
\end{equation}
where the probability amplitudes $\alpha_i \in \mathbb{C}$ satisfy
\begin{equation}
\displaystyle\sum_{i\in{\{0, 1\}}^{\it n}} \abs{\alpha_{i}}^{2}=1.
\label{eq2}
\end{equation}
Therefore, a quantum state of $n$ qubits can be alternatively represented as a $2^n$-dimensional state vector ${[\alpha_{0},...,\;\alpha_{2^{{\it n}-1}}]}^{T}$.

For initial state preparation, a quantum circuit simulator needs to construct a state vector representing some specified initial quantum state.

\subsubsection{State Evolution via Quantum Circuit Operation}
The state of a quantum system can be updated by the application of quantum operations (or so-called quantum gates).
The functionality of a quantum operation applied on $n$ qubits can be described by a
$2^{n} \times 2^{n}$ dimensional \emph{unitary matrix} $U$ (satisfying $U^{-1} = U^\dag$, that is, its inverse matrix equals its Hermitian adjoint).
The quantum gates considered in this work are listed in Table~\ref{tb:gate}.

\iffalse
\textcolor{blue}{
\begin{example}
The symbols and unitary matrices of quantum operations supported in our simulator are shown in Table~\ref{tb:gate}. The followings describe some examples: the Hadamard gate H sets the applied qubit into a superposition; the Phase gate S shifts the phase of the applied qubit by $\imath$; the controlled-NOT gate CNOT is a 2-qubit gate which complements the target (denoted by ${\oplus}$) when the control (denoted by ${\bullet}$) is 1; the Toffoli gate, which allows multiple controls, is the generalized version of the controlled-NOT gate; the Fredkin gate is a multi-qubit gate which swaps its two targets (denoted by $\times$) when the control(s) is (are) 1.
\end{example}
}
\fi

\begin{table}[]
\centering
\caption{Quantum gates supported in this work.}
\begin{tabular}{lcc}
\toprule[.1em]
Gate & Symbol & Matrix \\
\midrule[.1em]
Pauli-X (X) & \includegraphics[scale=0.2, valign=c]{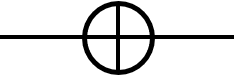} &  $\begin{bmatrix}
0 & 1 \\
1 & 0
\end{bmatrix}$\\
Pauli-Y (Y) & \includegraphics[scale=0.5, valign=c]{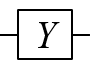} &
$\begin{bmatrix}
0 & -\imath \\
\imath & 0
\end{bmatrix}$\\
Pauli-Z (Z) & \includegraphics[scale=0.5, valign=c]{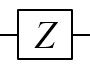} &
$\begin{bmatrix}
1 & 0 \\
0 & -1
\end{bmatrix}$\\
Hadamard (H) &\includegraphics[scale=0.5, valign=c]{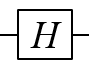} & $\frac{1}{\sqrt{2}}
\begin{bmatrix}
1 & 1 \\
1 & -1
\end{bmatrix}$ \\
Phase (S) & \includegraphics[scale=0.5, valign=c]{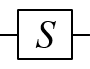} &
$\begin{bmatrix}
1 & 0 \\
0 & \imath
\end{bmatrix}$\\
T & \includegraphics[scale=0.5, valign=c]{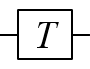} &
$\begin{bmatrix}
1 & 0 \\
0 & e^{\imath{\pi}/4}
\end{bmatrix}$\\
&&\\
Rx($\frac{\pi}{2}$) & \includegraphics[scale=0.25, valign=c]{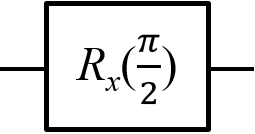} &
$\frac{1}{\sqrt{2}}
\begin{bmatrix}
1 & -\imath \\
-\imath & 1
\end{bmatrix}$\\
&&\\
Ry($\frac{\pi}{2}$) & \includegraphics[scale=0.25, valign=c]{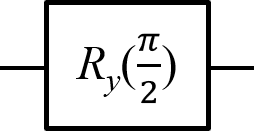} &
$\frac{1}{\sqrt{2}}
\begin{bmatrix}
1 & -1 \\
1 & 1
\end{bmatrix}$\\
&&\\
Controlled-NOT (CNOT) & \includegraphics[scale=0.2, valign=c]{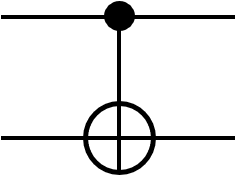} & $\scalemath{0.8}{
\begin{bmatrix}
1 & 0 & 0 & 0 \\
0 & 1 & 0 & 0 \\
0 & 0 & 0 & 1 \\
0 & 0 & 1 & 0
\end{bmatrix}
}$ \\
&&\\
Controlled-Z (CZ) & \includegraphics[scale=0.5, valign=c]{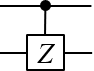} &
$\scalemath{0.8}{
\begin{bmatrix}
1 & 0 & 0 & 0 \\
0 & 1 & 0 & 0 \\
0 & 0 & 1 & 0 \\
0 & 0 & 0 & -1
\end{bmatrix}
}$\\
&&\\
Toffoli & \includegraphics[scale=0.2, valign=c]{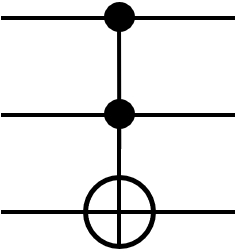} &
$\scalemath{0.6}{
\begin{bmatrix}
1 & 0 & 0 & 0 & 0 & 0 & 0 & 0 \\
0 & 1 & 0 & 0 & 0 & 0 & 0 & 0 \\
0 & 0 & 1 & 0 & 0 & 0 & 0 & 0 \\
0 & 0 & 0 & 1 & 0 & 0 & 0 & 0 \\
0 & 0 & 0 & 0 & 1 & 0 & 0 & 0 \\
0 & 0 & 0 & 0 & 0 & 1 & 0 & 0 \\
0 & 0 & 0 & 0 & 0 & 0 & 0 & 1 \\
0 & 0 & 0 & 0 & 0 & 0 & 1 & 0
\end{bmatrix}
}$\\
&&\\
Fredkin & \includegraphics[scale=0.2, valign=c]{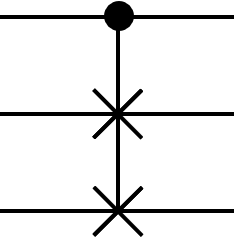} &
$\scalemath{0.6}{
\begin{bmatrix}
1 & 0 & 0 & 0 & 0 & 0 & 0 & 0 \\
0 & 1 & 0 & 0 & 0 & 0 & 0 & 0 \\
0 & 0 & 1 & 0 & 0 & 0 & 0 & 0 \\
0 & 0 & 0 & 1 & 0 & 0 & 0 & 0 \\
0 & 0 & 0 & 0 & 1 & 0 & 0 & 0 \\
0 & 0 & 0 & 0 & 0 & 0 & 1 & 0 \\
0 & 0 & 0 & 0 & 0 & 1 & 0 & 0 \\
0 & 0 & 0 & 0 & 0 & 0 & 0 & 1
\end{bmatrix}
}$\\
\bottomrule[.1em]
\end{tabular}
\label{tb:gate}
\end{table}

Consequently the state of a quantum system can be updated by multiplying a state vector with a unitary matrix.
For a quantum circuit with $m$ gates corresponding to unitary operators $M_1, \ldots, M_m$ in order, let $v_0$ be the initial input state and $v_m$ be the final output state of the circuit.
Then
\begin{equation}
v_m = M_{m} \times M_{m-1} \times {\cdots} \times M_1 \times v_0.
\label{e3}
\end{equation}
%The subscrpit indicates the order of the application. ${\it v_0}$ and ${\it v_r}$ are the initial and the resulting state vector, respectively.

\subsubsection{Qubit Measurement}
After all quantum gates are applied, we may need to measure some qubits to get their state outcome and determine its probability.
%obtain the final numerical result.
In quantum mechanics, measuring a qubit makes its state collapse (with superposition being destroyed) into an eigenstate with respect to the measurement basis.
The probability of an eigenstate being observed is determined by its corresponding probability amplitude.
Specifically, the probability of qubit $q$ being collapsed to $\ket{0}$ (similarly $\ket{1}$) can be calculated by
\begin{equation}
\Pr[q = \ket{0}] = \displaystyle \sum_{i \in \{0,1\}^n \mathrm{\ with\ bit\ } q =0} \abs{\alpha_{i}}^{2}.
\label{eq4}
\end{equation}

Suppose after measuring qubit $q$, it collapses to state $\ket{0}$, say.
Then the quantum state of the $n$-qubit system will have 0 probability amplitudes for states $\ket{i}$ for $i \in \{0,1\}^n$ with bit $q$ being 1.
Consequently the probability amplitudes for the other states, i.e., $\ket{i}$ for $i \in \{0,1\}^n$ with bit $q$ being 0, are renormalized by the factor $1/\sqrt{\Pr[q = \ket{0}]}$.

Note that the measurement process can be repeated for some other qubits sequentially or simultaneously.

\iffalse
\textcolor{blue}{
\begin{example}
Fig.~\ref{fig1} shows the procedure of measuring $q_0$ of a state vector (the left one) of a 3-qubit quantum system. The probabilities of $q_0=\ket{0}$ and $q_0=\ket{1}$ can be computed by \eqref{eq4},
\begin{equation*}
P(q_0=\ket{0})=(\frac{1}{2})^2\times(\frac{\sqrt{2}}{2})^2=\frac{3}{4}\text{, and}
\end{equation*}
\begin{equation*}
P(q_0=\ket{1})=(\frac{1}{2})^2=\frac{1}{4}.
\end{equation*}
Assume we measure $q_0=\ket{0}$, then the state vector is updated to set the amplitudes with $q_i=\ket{1}$ to zero (the middle one). Finally, the state vector is multiplied by the normalized factor $1/\sqrt{3/4}=2/\sqrt{3}$ to meet the normalization constraint (the right one).
\end{example}
}

\begin{figure}[htbp]
\centerline{\includegraphics[width=\columnwidth]{fig/measurement.PNG}}
\caption{Example of measuring one qubit.}
\label{fig1}
\end{figure}
\fi

\section{BDD-Based Quantum Simulation} \label{sec:alg}
In this section, we describe the proposed BDD-based structure for representing state vectors, and provide methods to complete quantum simulation on this BDD-based structure, i.e., realize three actions discussed in Section~\ref{sec:prelim}.

\subsection{Algebraic Representation of Complex Values}
In this work, we employ the algebraic representation of complex values proposed in \cite{accuracy} to achieve \emph{accurate} quantum simulation without precision loss.
Essentially, any complex value (scalar) $\alpha$ that can be represented exactly can be expressed as
\begin{equation}\label{eq:algebraic}
\alpha = \frac{1}{{\sqrt{2}}^k}(a{\omega}^3+b{\omega}^2+c{\omega}+d),
\end{equation}
where the coefficients $a, b, c, d, k \in \mathbb{Z}$, $\alpha\in\mathbb{C}$, and $\omega=e^{\imath{\pi}/4}$,
Therefore, in the context of quantum circuit simulation, when all the entries in the initial state vector and the unitary operators can be exactly represented, all the complex values result from the matrix-vector multiplication can be exactly represented, too.

The representation is appealing as we only need to maintain five integers to represent a complex number.
By a simple counting argument that $\mathbb{Z}^5$ is countable while $\mathbb{C}$ is uncountable, clearly not every complex number can be compactly represented in this algebraic form.
However as there are universal gate sets, such as the Clifford+T set, for quantum computing whose entries are all exactly representable, quantum circuit simulation under the algebraic representation can be generally done without loss of generality.
It is because the universality of a gate set allows any unitary operator to be approximated using the gates from the set within any desired precision.
Hence as long as the initial state can be exactly represented, quantum circuit simulation can be achieved without precision loss.

%This means instead of storing floating point numbers for a complex value, this representation stores five integers and manipulates the corresponding complex value.
%According to \cite{accuracy}, all quantum operations which can be realized exactly by the universal quantum gate library Clifford+T\cite{b4} can be represented perfectly. More precisely, the matrices of these quantum operations can be represented without any information loss by using \eqref{eq:algebraic}.

The next question we should address is how to efficiently maintain and manipulate the integers for quantum circuit simulation.

%We use \eqref{eq:algebraic} in our simulation approach not only to minimize the information loss but also to provide our proposed BDD-based structure with the easier manipulation of Boolean operations  (i.e. operations on integers instead of floating point numbers).

\subsection{Bit-Slicing State Vectors with BDDs}\label{sec:slice}
Given a state vector $\ket{\psi}$ of an $n$-qubit quantum system with its entries represented algebraically in the form of Eq.~\eqref{eq:algebraic}, we exploit BDDs for representation as follows.

\begin{figure}[t]
\centerline{\includegraphics[width=\columnwidth]{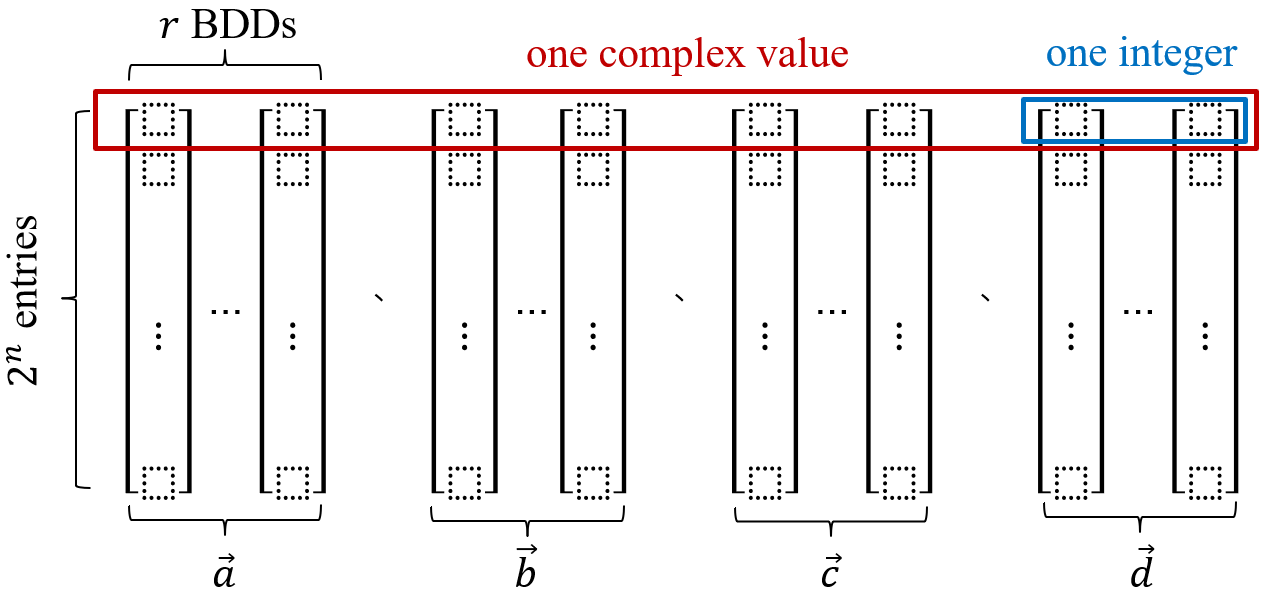}}
\caption{Bit-slicing algebraic numbers with BDDs.}
\label{fig2}
\end{figure}

By Eq.~\eqref{eq:algebraic}, the state vector $\ket{\psi}$ (with $2^n$ entries of complex values) is described by an integer scalar $k$ and four vectors $\vec{a} = [a_0, \ldots, a_{2^{n-1}}]^T, \vec{b} = [b_0, \ldots, b_{2^{n-1}}]^T, \vec{c} = [c_0, \ldots, c_{2^{n-1}}]^T, \vec{d} = [d_0, \ldots, d_{2^{n-1}}]^T$, each with $2^n$ entries of integer values.
Hence the probability amplitude $\alpha_i$ of basis state $\ket{i}$ in $\ket{\psi}$ equals $\frac{1}{\sqrt{2}^k}(a_i \omega^3 + b_i \omega^2 + c_i \omega + d_i)$ for $i \in \{0,1\}^n$.
Let $a_i, b_i, c_i, d_i$ be $r$-bit integers ($r$ can be adjusted on-the-fly as large as enough in our implementation).
We represent the $j^\mathrm{th}$ bit, for $j=1, \ldots, r$, of each of the vectors $\vec{a}, \vec{b}, \vec{c}, \vec{d}$ with a BDD as illustrated in Fig.~\ref{fig2}.
Thereby each BDD is a function over the $n$ qubit variables so that the truth table of the function corresponds to the corresponding a bit vector of $2^n$ entries.
On the other hand, because scalar $k$ is shared among all $2^n$ entries of $\vec{a}, \vec{b}, \vec{c}, \vec{d}$, it is stored separately.
Overall we use $4r$ BDDs over $n$ variables to represent an $n$-qubit state vector.
In the sequel, we denote the BDDs of the $i^\mathrm{th}$ bit of $\vec{a}$, $\vec{b}$, $\vec{c}$, and $\vec{d}$ as $F^{ai}$, $F^{bi}$, $F^{ci}$, and $F^{di}$, respectively.

%\begin{theorem}
%Given a bit vector $[w_0,\cdots, w_{2^{n-1}}]^{T}$ with $w_{i}\in\mathbb{B}$, it can be represented by a BDD with $n$ Boolean variables $x_0,\cdots,x_{n-1}$. The value of the BDD when the binary $x_0\cdots x_{n-1}$ equals to the integer $i$ is $w_{i}$.
%\end{theorem}

\iffalse
\begin{example}
Given a 2-qubit state vector with the integer vector of $d$ being $[2, 0, 2, 3]^T$, it can be decomposed into $r$ bit vectors.
The bit vector for LSB (i.e. $F^{d0}$) is $[0, 0, 0, 1]^T$, which can be represented as the BDD shown in Fig.~\ref{fig3}.
\end{example}

\begin{figure}[htbp]
\centerline{\includegraphics[scale=0.4]{fig/BDD_for_bit_vector.PNG}}
\caption{Example of measuring one qubit.}
\label{fig3}
\end{figure}
\fi

Note that we require no algebraic representation for unitary matrices as we replace matrix-vector multiplication with Boolean formula manipulation over state-vector BDDs as to be explained in Section~\ref{sec:formula}.

\subsection{Initial State Construction}
A typical, if not every, quantum computation algorithm prepares its initial state in one of the basis states specified \textit{a priori}.
In this case, all BDDs correspond to constant 0 (logical false) except for $F^{d0}$.
Given an $n$-qubit quantum circuit with an initial state $\ket{i} = \ket{b_0 \ldots b_{n-1}}$ for some $i \in \{0,1\}^n$ and $b_0,\ldots,b_{n-1} \in \{0,1\}$, then
\begin{equation}
F^{d0} = \bigwedge_{j=0}^{n-1} l_j,
\;\;
\mbox{ for }
l_j=\left\{
\begin{aligned}
\overbar{q_j},\; \mbox{ if } b_j=0\\
q_j,\; \mbox{ if } b_j=1
\end{aligned}
\right.,
\end{equation}
where $q_j$ is the Boolean variable of the formula/BDD under construction corresponding to the $j^\mathrm{th}$ qubit of the quantum circuit.

\iffalse
\begin{example}
{\it Given a 5-qubit quantum system and assume the initial basis state set by users is $\ket{q_0q_1q_2q_3q_4}=\ket{01001}$, then
\begin{equation*}
F^{ai}=F^{bi}=F^{ci}=0\;\;\text{for}\;\;0\leq i \leq r-1\text{, and}
\end{equation*}
\begin{equation*}
F^{di}=\left\{
\begin{aligned}
\overbar{{x_0}}x_1\overbar{{x_2}}\overbar{{x_3}}x_4,\;\;i=0\\
0,\;\;1\leq i \leq r-1
\end{aligned}
\right..
\end{equation*}}
\end{example}
% \begin{equation}
% F^{ai}=F^{bi}=F^{ci}=0\text{, and}
% \end{equation}
% \begin{equation}
% F^{di}=\left\{
% \begin{aligned}
% g_0\cdots g_{n-2}g_{n-1},\;j=0\\
% 0,\;1\leq j \leq r-1
% \end{aligned}
% \right.
% ,\;
% g_i=\left\{
% \begin{aligned}
% \overbar{x_i},\;\ket{q_i}=\ket{0}\\
% x_i,\;\ket{q_i}=\ket{1}
% \end{aligned}
% \right.
% \end{equation}
\fi

\subsection{Unitary State Evolution via Formula Manipulation}
\label{sec:formula}
Under the bit-sliced state vector representation using BDDs as detailed in Section~\ref{sec:slice}, we show how to update a state vector when an quantum gate in Table~\ref{tb:gate} is applied.
Essentially, the new state is the result of the matrix-vector multiplication.

We note that the gates in Table~\ref{tb:gate} form a library general enough to achieve universal quantum computation as it is a superset of the Clifford+T \cite{b4} as well as the Toffoli+Hadamard \cite{b5} universal gate sets.
%Due to this universality, any quantum operations can be simulated by our simulator up to an small error.

To avoid vector-matrix multiplication but achieve the same computation, we examine the effect of every considered quantum gate and characterize its corresponding Boolean formulas for updating the bit-sliced algebraic parameters $(\vec{a},\vec{b},\vec{c},\vec{d},k)$ stated in Section~\ref{sec:slice}.
The obtained update rules for $F^{ai}, F^{bi}, F^{ci}, F^{di}$ are summarized in Table~\ref{tb:formula}.
For brevity, below we only detail the update rule derivation of H-gate while leaving others as exercises for interested readers to verify.

\begin{table*}[]
\caption{Boolean formulas for quantum state evolution of the supported gate set.}
\begin{tabular}{lllll}
\toprule[.1em]
\multirow{2}{*}{Gate} & \multicolumn{4}{c}{Boolean formulas} \\
                           & Update $F^{ai}$  & Update $F^{bi}$  & Update $F^{ci}$  & Update $F^{di}$ \\
\midrule[.1em]
X                       & ${\hat{F}}^{ai} = q_t  F^{ai}|_{\overbar{q_t}}\vee   \overbar{q_t}  F^{ai}|_{q_t}.$   &\multicolumn{3}{c}{same as $F^{ai}$ except renaming}    \\
\midrule[.1em]
\multirow{5}{*}{Y}   &$G^{ai}=q_t  F^{ci}|_{\overbar{q_t}}\vee \overbar{q_t}  F^{ci}|_{q_t},$&$G^{bi}=q_t  F^{di}|_{\overbar{q_t}}\vee \overbar{q_t}  F^{di}|_{q_t},$&$G^{ci}=q_t  F^{ai}|_{\overbar{q_t}}\vee \overbar{q_t}  F^{ai}|_{q_t},$&$G^{di}=q_t  F^{bi}|_{\overbar{q_t}}\vee \overbar{q_t}  F^{bi}|_{q_t},$\\
                           &$D^{ai}=q_t  G^{ai}\vee \overbar{q_t}  \overbar{G^{ai}},$&$D^{bi}=q_t  G^{bi}\vee \overbar{q_t}  \overbar{G^{bi}},$&$D^{ci}=q_t  \overbar{G^{ci}}\vee \overbar{q_t}  G^{ci},$&$D^{di}=q_t  \overbar{G^{di}}\vee \overbar{q_t}  G^{di},$\\ &$C^{a0}=\overbar{q_t},$&$C^{b0}=\overbar{q_t},$&$C^{c0}=q_t,$&$C^{d0}=q_t,$\\
                           &$C^{a(i+1)} = \mathrm{Car}(D^{ai}, 0, C^{ai}),$&$C^{b(i+1)} = \mathrm{Car}(D^{bi}, 0, C^{bi}),$&$C^{c(i+1)} = \mathrm{Car}(D^{ci}, 0, C^{ci}),$&$C^{d(i+1)} = \mathrm{Car}(D^{di}, 0, C^{di}),$\\
                           &${\hat{F}}^{ai}= \mathrm{Sum}(D^{ai}, 0, C^{ai}).$&${\hat{F}}^{bi}= \mathrm{Sum}(D^{bi}, 0, C^{bi}).$&${\hat{F}}^{ci}= \mathrm{Sum}(D^{ci}, 0, C^{ci}).$&${\hat{F}}^{di}= \mathrm{Sum}(D^{di}, 0, C^{di}).$\\
\midrule[.1em]
\multirow{4}{*}{Z} &   $G^{ai}=\overbar{q_t}  F^{ai}\vee q_t \overbar{F^{ai}},$ &\multicolumn{3}{c}{\multirow{4}{*}{same as $F^{ai}$ except renaming}} \\
                            &$C^{a0} = q_t,$ & \multicolumn{3}{c}{} \\
                              & $C^{a(i+1)} = \mathrm{Car}(G^{ai}, 0, C^{ai}),$  & \multicolumn{3}{c}{} \\
                              & ${\hat{F}}^{ai}= \mathrm{Sum}(G^{ai}, 0, C^{ai}).$  & \multicolumn{3}{c}{}  \\
\midrule[.1em]
\multirow{5}{*}{H}     & $G^{ai} = F^{ai}|_{\overbar{q_t}},$ &\multicolumn{3}{c}{\multirow{5}{*}{same as $F^{ai}$ except renaming}} \\
                              & $D^{ai} = \overbar{q_t}  F^{ai}|_{q_t} \vee  q_t  \overbar{F^{ai}},$     & \multicolumn{3}{c}{}            \\
                              &$C^{a0} = q_t,$\\
                              & $C^{a(i+1)} = \mathrm{Car}(G^{ai}, D^{ai}, C^{ai}),$  & \multicolumn{3}{c}{}  \\
                              & ${\hat{F}}^{ai}= \mathrm{Sum}(G^{ai}, D^{ai}, C^{ai}).$  & \multicolumn{3}{c}{}  \\
\midrule[.1em]
\multirow{4}{*}{S}         &   ${\hat{F}}^{ai}=\overbar{q_t}  F^{ai}\vee q_t  F^{ci}.$ & ${\hat{F}}^{bi}=\overbar{q_t}  F^{bi}\vee q_t  F^{di}.$&$G^{ci}=\overbar{q_t}  F^{ci}\vee q_t \overbar{F^{ai}},$&$G^{di}=\overbar{q_t}  F^{di}\vee q_t \overbar{F^{bi}},$\\
                           &         &         & $C^{c0} = q_t,$ & $C^{d0} = q_t,$ \\
                           &         &         & $C^{c(i+1)} = \mathrm{Car}(G^{ci}, 0, C^{ci}),$ & $C^{d(i+1)} = \mathrm{Car}(G^{di}, 0, C^{di}),$ \\
                           &         &         & ${\hat{F}}^{ci}= \mathrm{Sum}(G^{ci}, 0, C^{ci}).$ & ${\hat{F}}^{di}= \mathrm{Sum}(G^{di}, 0, C^{di}).$ \\
\midrule[.1em]
\multirow{4}{*}{T}         &   ${\hat{F}}^{ai}=\overbar{q_t}  F^{ai}\vee q_t  F^{bi}.$ & ${\hat{F}}^{bi}=\overbar{q_t}  F^{bi}\vee q_t  F^{ci}.$&${\hat{F}}^{ci}=\overbar{q_t}  F^{ci}\vee q_t  F^{di}.$&$G^{di}=\overbar{q_t}  F^{di}\vee q_t \overbar{F^{ai}},$\\
                           &         &         &         & $C^{d0} = q_t,$ \\
                           &         &         &         & $C^{d(i+1)} = \mathrm{Car}(G^{di}, 0, C^{di}),$ \\
                           &         &         &         & ${\hat{F}}^{di}= \mathrm{Sum}(G^{di}, 0, C^{di}).$ \\
\midrule[.1em]
\multirow{5}{*}{Rx($\frac{\pi}{2}$)}        &$D^{ai}=q_t  F^{ci}|_{\overbar{q_t}}\vee \overbar{q_t}  F^{ci}|_{q_t},$&$D^{bi}=q_t  F^{di}|_{\overbar{q_t}}\vee \overbar{q_t}  F^{di}|_{q_t},$&$D^{ci}=q_t  F^{ai}|_{\overbar{q_t}}\vee \overbar{q_t}  F^{ai}|_{q_t},$&$D^{di}=q_t  F^{bi}|_{\overbar{q_t}}\vee \overbar{q_t}  F^{bi}|_{q_t},$\\
                           &$C^{a0}=1,$&$C^{b0}=1,$&$C^{c0}=0,$&$C^{d0}=0,$\\
                           &$C^{a(i+1)} = \mathrm{Car}(F^{ai}, \overbar{D^{ai}}, C^{ai}),$&$C^{b(i+1)} = \mathrm{Car}(F^{bi}, \overbar{D^{bi}}, C^{bi}),$&$C^{c(i+1)} = \mathrm{Car}(F^{ci}, D^{ci}, C^{ci}),$&$C^{d(i+1)} = \mathrm{Car}(F^{di}, D^{di}, C^{di}),$\\
                           &${\hat{F}}^{ai}= \mathrm{Sum}(F^{ai}, \overbar{D^{ai}}, C^{ai}).$&${\hat{F}}^{bi}= \mathrm{Sum}(F^{bi}, \overbar{D^{bi}}, C^{bi}).$&${\hat{F}}^{ci}= \mathrm{Sum}(F^{ci}, D^{ci}, C^{ci}).$&${\hat{F}}^{di}= \mathrm{Sum}(F^{di}, D^{di}, C^{di}).$\\
\midrule[.1em]
\multirow{5}{*}{Ry($\frac{\pi}{2}$)} &$G^{ai} = F^{ai}|_{\overbar{q_t}},$ &\multicolumn{3}{c}{\multirow{5}{*}{same as $F^{ai}$ except renaming}}\\
                           &$D^{ai} = q_t  F^{ai} \vee  \overbar{q_t}  \overbar{F^{ai}|_{q_t}},$ & \multicolumn{3}{c}{} \\
                           &$C^{a0} = \overbar{q_t},$ & \multicolumn{3}{c}{} \\
                              & $C^{a(i+1)} = \mathrm{Car}(G^{ai}, D^{ai}, C^{ai}),$ & \multicolumn{3}{c}{}   \\
                              & ${\hat{F}}^{ai}= \mathrm{Sum}(G^{ai}, D^{ai}, C^{ai}).$  & \multicolumn{3}{c}{}  \\
\midrule[.1em]
\multirow{2}{*}{CNOT}  & ${\hat{F}}^{ai} = \overbar{q_c}  F^{ai}\vee q_c  q_t  F^{ai}|_{q_c \overbar{q_t}}$ &\multicolumn{3}{c}{\multirow{2}{*}{same as $F^{ai}$ except renaming}} \\
                       & $ \qquad\quad \vee q_c  \overbar{q_t}  F^{ai}|_{q_c q_t}.$ & \multicolumn{3}{c}{} \\
\midrule[.1em]
\multirow{4}{*}{CZ} &   $G^{ai}=\overbar{q_c q_t}  F^{ai}\vee q_c q_t \overbar{F^{ai}},$  &\multicolumn{3}{c}{\multirow{4}{*}{same as $F^{ai}$ except renaming}}    \\
                            &$C^{a0} = q_c q_t,$ & \multicolumn{3}{c}{} \\
                              & $C^{a(i+1)} = \mathrm{Car}(G^{ai}, 0, C^{ai}),$  & \multicolumn{3}{c}{}  \\
                              & ${\hat{F}}^{ai}= \mathrm{Sum}(G^{ai}, 0, C^{ai}).$  & \multicolumn{3}{c}{}  \\
\midrule[.1em]
\multirow{2}{*}{Toffoli}    & ${\hat{F}}^{ai} = \overbar{Q_c}  F^{ai}\vee Q_c  q_t  F^{ai}|_{Q_c \overbar{q_t}}$ &\multicolumn{3}{c}{\multirow{2}{*}{same as $F^{ai}$ except renaming}} \\
                            & $\qquad\quad \vee Q_c  \overbar{q_t}  F^{ai}|_{Q_c q_t}.$ & \multicolumn{3}{c}{} \\
\midrule[.1em]
\multirow{3}{*}{Fredkin}    &  ${\hat{F}}^{ai} = \overbar{Q_c(q_{t}\oplus {q_{t^{'}}})}F^{ai}$ &\multicolumn{3}{c}{\multirow{3}{*}{same as $F^{ai}$ except renaming}} \\
                            & $\qquad\quad \vee Q_c q_{t} \overbar{q_{t^{'}}} F^{ai}|_{Q_c \overbar{q_{t}} q_{t^{'}}}$ & \multicolumn{3}{c}{} \\
                            &$\qquad\quad \vee Q_c \overbar{q_{t}} q_{t^{'}} F^{ai}|_{Q_c q_{t} \overbar{q_{t^{'}}}}.$   & \multicolumn{3}{c}{}  \\
\bottomrule[.1em]
\end{tabular}
\label{tb:formula}
\end{table*}

For simplicity, consider a 2-qubit quantum system with the state vector $[\alpha_0, \alpha_1, \alpha_2, \alpha_3]^T$, where $\alpha_0$, $\alpha_1$, $\alpha_2$, and $\alpha_3$ are the probability amplitudes of basis states $\ket{q_0 q_1}=\ket{00}$, $\ket{01}$, $\ket{10}$, and $\ket{11}$, respectively.
Without loss of generality, assume a Hadamard gate is applied on qubit $q_0$.
The unitary matrix for the 2-qubit system is then obtained by the Kronecker product
\begin{equation*}
H\otimes I=\frac{1}{\sqrt{2}}
\begin{bmatrix}
1 & 1 \\
1 & -1
\end{bmatrix}
\otimes
\begin{bmatrix}
1 & 0 \\
0 & 1
\end{bmatrix}
=\frac{1}{\sqrt{2}}
\begin{bmatrix}
1 & 0 & 1 & 0 \\
0 & 1 & 0 & 1 \\
1 & 0 & -1 & 0 \\
0 & 1 & 0 & -1
\end{bmatrix}.
\end{equation*}
With the scaling factor $\frac{1}{\sqrt{2}}$ being omitted (which contributes to the increment of parameter $k$ by 1), the state vector is updated to
\begin{equation*}
%\frac{1}{\sqrt{2}}
\begin{bmatrix}
1 & 0 & 1 & 0 \\
0 & 1 & 0 & 1 \\
1 & 0 & -1 & 0 \\
0 & 1 & 0 & -1
\end{bmatrix}
\begin{bmatrix}
\alpha_0 \\ \alpha_1 \\ \alpha_2 \\ \alpha_3
\end{bmatrix}
=%\frac{1}{\sqrt{2}}
\begin{bmatrix}
\alpha_0+\alpha_2 \\ \alpha_1+\alpha_3 \\ \alpha_0-\alpha_2 \\ \alpha_1-\alpha_3
\end{bmatrix}
=
%\frac{1}{\sqrt{2}}
\begin{bmatrix}
\alpha_0 \\ \alpha_1 \\ \alpha_0 \\ \alpha_1
\end{bmatrix}
+
\begin{bmatrix}
\alpha_2 \\ \alpha_3 \\ -\alpha_2 \\ -\alpha_3
\end{bmatrix}.
\end{equation*}
\iffalse
and can be rewritten as the sum of two vectors:
\begin{equation*}
\frac{1}{\sqrt{2}}
\begin{bmatrix}
w_0+w_2 \\ w_1+w_3 \\ w_0-w_2 \\ w_1-w_3
\end{bmatrix}
=
\frac{1}{\sqrt{2}}
(
\begin{bmatrix}
w_0 \\ w_1 \\ w_0 \\ w_1
\end{bmatrix}
+
\begin{bmatrix}
w_2 \\ w_3 \\ -w_2 \\ -w_3
\end{bmatrix}
)
\end{equation*}
\fi
%It is obvious that these two vectors can be derived from the input state vector by the rearrangement of its entries.
%Consequently, we need to generalize the Boolean formulas
%\begin{enumerate}[label=(\roman*)]
%\item to derive two component vectors from the input state vector, and
%\item to sum two component vectors.
%\end{enumerate}
%The generalized Boolean formulas for Hadamard gate, to update the integer vector of $a$, are provided in theorem \ref{th_H}.
%Note that, to distinguish between input and output BDDs of the state vector, we add carets on top of upper-case letters to denote the output BDDs in the sequel (e.g. $\hat{F}^{ai}$).
The summation of the above two vectors forms the basis of updating $F^{ai}, F^{bi}, F^{ci}, F^{di}$ for the Hadamard gate.

For a general $n$-qubit quantum system, the following proposition specifies the formulas that update $F^{ai}$ (similarly $F^{bi}, F^{ci}, F^{di}$) of the original state vector to $\hat{F}^{ai}$ of the new state vector.
\begin{proposition}
\label{th_H}
For an $n$-qubit quantum system with its state vector being algebraically represented with $F^{ai}$ (along with $F^{b^i}, F^{c^i}, F^{d^i}$), let Hadamard gate be applied on an arbitrary qubit $q_t$.
Then the new state vector is specified by $\hat{F}^{ai}$ (along with $\hat{F}^{b^i}, \hat{F}^{c^i}, \hat{F}^{d^i}$ derivable similarly) as follows.
\begin{eqnarray}
G^{ai} & = & F^{ai}|_{\overbar{q_t}}, \label{eq8}\\
D^{ai} & = & \overbar{q_t} F^{ai}|_{q_t} \vee q_t \overbar{F^{ai}},\label{eq9}\\
C^{a0} & = & q_t,\label{eq10}\\
C^{a(i+1)} & = & G^{ai}D^{ai} \vee (G^{ai} \vee D^{ai})C^{ai},\label{eq11}\\
{\hat{F}}^{ai} & = & G^{ai} \oplus D^{ai} \oplus C^{ai},\label{eq12}
\end{eqnarray}
where $i=0, \ldots, r-1$ for $r$ be the integer size of algebraic parameters $(\vec{a},\vec{b},\vec{c},\vec{d})$. Moreover, the $k$-value increases by 1.
\end{proposition}

To see the correctness (with the help of the above 2-qubit example), observe that Eq.~\eqref{eq8} and Eq.~\eqref{eq9} derive the BDDs of two component vectors ($[\alpha_0,\alpha_1,\alpha_0,\alpha_1]^T$ and $[\alpha_2,\alpha_3,-\alpha_2,-\alpha_3]^T$, respectively).
Moreover, Eq.~\eqref{eq10}, Eq.~\eqref{eq11} and Eq.~\eqref{eq12} together fulfill the function of an adder summing the two component vectors bitwisely.
It is worth noting that since we use 2's complement to represent integers, we set $C^{a0} = q_t$ in Eq.~\eqref{eq10} as the initial carry-in of the adder function to realize the "plus one" action for the negated entries of the second component vector (in $[\alpha_2,\alpha_3,-\alpha_2,-\alpha_3]^T$ the last two entries are complemented in Eq.~\eqref{eq9} by the second term, i.e., $q_t \overline{F^{ai}}$, and their "plus one" actions are realized by the initial carry-in setting).
On the other hand, the Hadamard gate increases the $k$-value by 1 due to the $\frac{1}{\sqrt{2}}$ scaling factor.

For other quantum gates, their $F^{ai}$, $F^{bi}$, $F^{ci}$, and $F^{di}$ formulas are obtained and summarized in Table~\ref{tb:formula}, where $q_c$ is the control bit, $Q_c$ is the conjunction of all control bits of the Toffoli and Fredkin gate, $q_t$ is the target bit, $q_{t^{'}}$ is the second target for Fredkin gates, and $i=0, \ldots, r-1$.
Also in the table, for brevity we define
\begin{eqnarray*}
\mathrm{Car}(A, B, C) & \triangleq & A B \vee (A \vee B) C, \\
\mathrm{Sum}(A, B, C) & \triangleq & A \oplus B \oplus C,
\end{eqnarray*}
to denote the \emph{carry} and \emph{sum} operations over formulas $A,B,C$.
We note that the formulas of the Toffoli gate in the table work for a general Toffoli gate of an arbitrary number of control bits.
Note also that quantum gates X, Z, H, Ry($\frac{\pi}{2}$), CNOT, CZ, Toffoli, and Fredkin involve no imaginary parts and thus their $F^{ai}$, $F^{bi}$, $F^{ci}$, and $F^{di}$ formulas are mutually independent.
In contrast, quantum gates Y, S, T, and Rx($\frac{\pi}{2}$) involve imaginary parts and cause phase shifts making their $F^{ai}$, $F^{bi}$, $F^{ci}$, and $F^{di}$ formulas mutually dependent.
About the algebraic parameter $k$, its value remains the same for all the quantum gates except for being incremented by 1 for Hadamard, Rx($\pi/2$), and Ry($\pi/2$) gates due to their $\frac{1}{\sqrt{2}}$ scaling factors.

\iffalse
Definition \ref{adder} writes the Boolean formulas of a full adder, such as Equation \eqref{eq11} and \eqref{eq12}, as functions in the remaining of this paper.
\begin{definition}
\label{adder}
{\it Given Boolean functions $A$, $B$, and $C_{in}$, the behavior of functions
\begin{equation*}
C_{out} = Car(A, B, C_{in}),
\end{equation*}
\begin{equation*}
P= Sum(A, B, C_{in})
\end{equation*}
are equal to the Boolean formulas
\begin{equation*}
C_{out} = A B \vee (A \vee B) C_{in},
\end{equation*}
\begin{equation*}
P=A \oplus B \oplus C_{in},
\end{equation*}
respectively.
}
\end{definition}
\fi

%\begin{theorem}\label{th_f}
%The Boolean formulas of the supported gates are given in Table~\ref{tb:formula}, where $q_c$ is the control, $Q_c$ is the conjunction of all controls, $q_t$ is the target, $q_g$ is the second target for Fredkin gates, and $i=0, \ldots, r-1$.
%When simulating a Hadamard, a Rx($\pi/2$), or a Ry($\pi/2$) gate, the integer $k$ in \eqref{eq:algebraic} is added by 1 due to the $1/{\sqrt{2}}$ factor of its unitary matrix.
%\end{theorem}

%Note that the Pauli-X and the CNOT gate are the Toffoli gates with no control (i.e. $q_c=1$) and with only one control, respectively, the Pauli-Z gate is the CZ gate with no control, and a S gate is equal to two successive T gates (i.e. $S=T^2$).

The correctness of Table~\ref{tb:formula} construction is asserted in the following theorem.
\begin{theorem}
The formulas $F^{ai}, F^{bi}, F^{ci}, F^{di}$ of Table~\ref{tb:formula} and the $k$ value update rule stated above correctly update the quantum state under the algebraic representation of Eq.~\eqref{eq:algebraic}.
\end{theorem}

%At the end of this subsection, we would like to point out that, in our simulator, 1) the initial state is one of the basis states, and 2) all entries of unitary matrices of the supported gates are exactly represented by \eqref{eq:algebraic}. Hence, entries of the state vector after applying each gate are also represented by \eqref{eq:algebraic} without any information loss.\footnote{The sum and product of two numbers in the form of \eqref{eq:algebraic} are also in the form of \eqref{eq:algebraic}.

We mention that in our implementation the integer bit size $r$ is augmented dynamically when necessary (i.e., overflow occurs).
Thereby with the accuracy guarantee of the algebraic representation, our simulation is exact (i.e., no precision loss).

\subsection{Measurement and Probability Calculation}

In the QMDD-based method \cite{zulehner2019advanced}, measurement and probability calculation can be done efficiently by traversing the QMDD.
For a qubit $q$ to be measured, placing it as the top variable of the QMDD makes $\Pr[q=\ket{0}]$ and $\Pr[q=\ket{0}]$ derivable in one QMDD traversal.

%A However, since bits of integers are stored separately in different BDDs in the proposed BDD-based structure, it is unlikely to use the same notion in \cite{zulehner2019advanced} directly on the BDD-based structure.
%To this end, we construct a big BDD $F$ to merge the information of each single BDD.
%This construction requires two (i.e. $\log_2 4$) extra variables to indicate four integers $a$, $b$, $c$, and $d$, and $\ceil{\log_2 r}$ extra variables to indicate the bit positions of the integers. Therefore, the number of variables of $F$ is $n+\ceil{\log_2 r}+2$. Definition \ref{def_F} shows the construction of $F$.

In our case, measurement and probability calculation are not as easy as the QMDD case because we do not have a monolithic BDD but rather $4r$ BDDs.
We adopt the hyperfunction construction \cite{hyper} of combining multiple BDDs into one monolithic BDD $F$ as follows.
Let
\begin{equation}
F = x_0 x_{1} F^{\vec{a}} \vee x_0 \overbar{x_{1}} F^{\vec{b}} \vee \overbar{x_0}x_{1} F^{\vec{c}} \vee \overbar{x_0}\overbar{x_{1}} F^{\vec{d}},
\end{equation}
for
\begin{eqnarray*}
F^{\vec{a}}  =  \bigvee\limits_{i=0}^{r-1} g_i F^{ai},
F^{\vec{b}}  =  \bigvee\limits_{i=0}^{r-1} g_i F^{bi},\\
F^{\vec{c}}  =  \bigvee\limits_{i=0}^{r-1} g_i F^{ci},
F^{\vec{d}}  =  \bigvee\limits_{i=0}^{r-1} g_i F^{di},
\end{eqnarray*}
with
\begin{equation*}
g_i =\left\{
\begin{aligned}
&\overbar{x_{2}}\overbar{x_{3}}\cdots\overbar{x_{\ceil{\log_2 r}+1}},&\;\;& \mbox{ if } i=0&\\
&x_{2}\overbar{x_{3}}\cdots\overbar{x_{\ceil{\log_2 r}+1}},&\;\;& \mbox{ if } i=1&\\
&\overbar{x_{2}}x_{3}\cdots\overbar{x_{\ceil{\log_2 r}+1}},&\;\;& \mbox{ if } i=2&\\
&&&\vdots&\\
&x_{2}x_{3}\cdots x_{\ceil{\log_2 r}+1},&\;\;& \mbox{ if } i= r-1&\\
\end{aligned}
\right.,
\end{equation*}
where $x_0$ to $x_{\ceil{\log_2 r}+1}$ are fresh new Boolean variables used for labeling/encoding the $4r$ BDDs.

\begin{figure}[t]
\centerline{\includegraphics[width=\columnwidth]{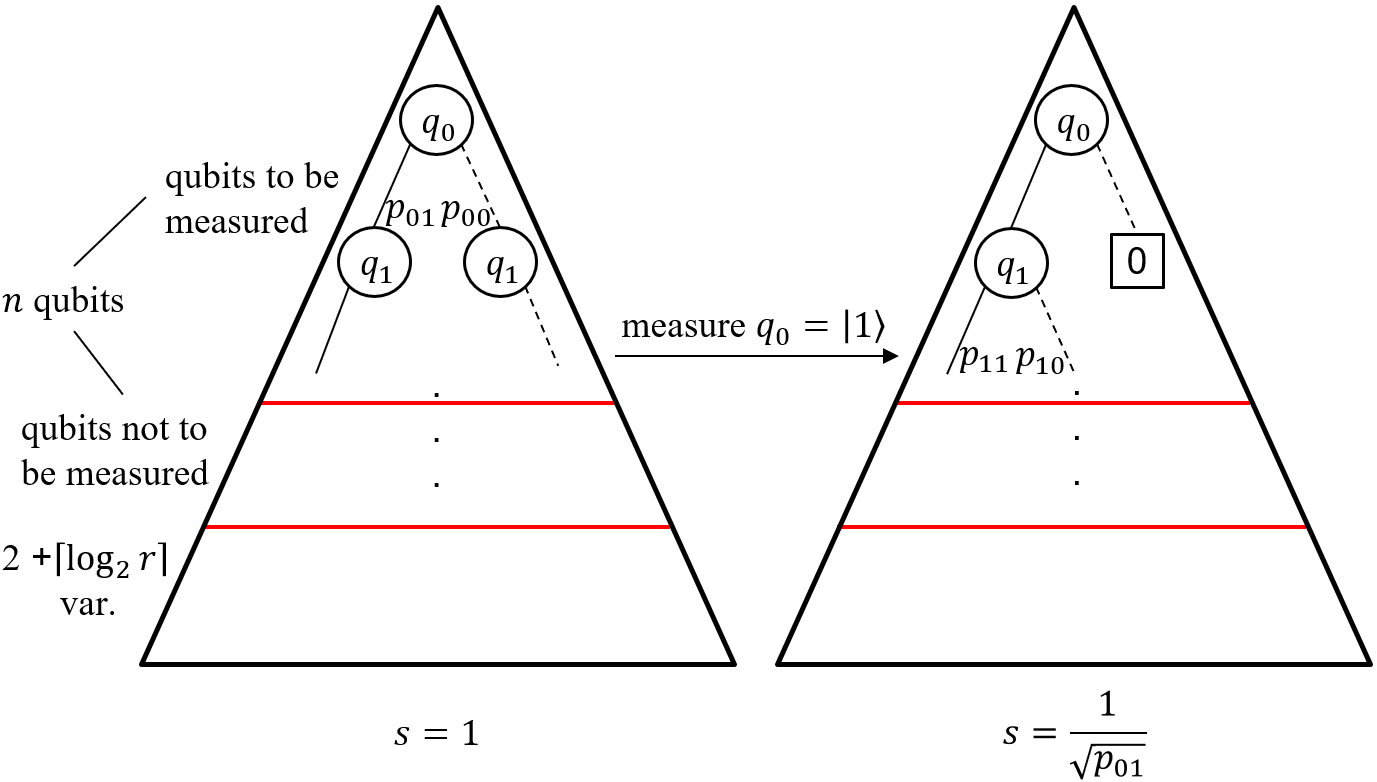}}
\caption{Monolithic BDD $F$ for measurement.}
\label{fig_big_BDD}
\end{figure}

Given a monolithic BDD $F$, the measurement procedure is conducted on $F$ as illustrated in Fig.~\ref{fig_big_BDD}, where the qubit variables $q_0, \ldots, q_{n-1}$ are ordered above the encoding variables $x_0, x_1$, which are followed by variables $x_2 \ldots, x_{\lceil \log_2 r \rceil+1}$ variables, and $p_{ij}$ denotes
the probability $\Pr[q_i=\ket{j}]$ for $j\in\{0,1\}$.
For probability calculation of measurement, we compute the accumulated probabilities of the nodes at the top $n$ levels recursively, and record them by a hash map.
(The accumulated probability of a node is the sum of the probabilities of its left and right children.)
If the recursive procedure reaches the $n^\mathrm{th}$ level of $F$ (counting from 0), the four algebraic integers $a$, $b$, $c$, and $d$ can be decoded by paths $x_0x_1$, $x_0\overbar{x_1}$, $\overbar{x_0}x_1$, and $\overbar{x_0}\overbar{x_1}$, respectively, to obtain
%\textcolor{red}{
values of the bit positions,
%?????????????????????},
and the corresponding probability amplitude $\alpha$ can be computed.

As mentioned in Section~\ref{sec:qcb}, all probability amplitudes should be multiplied with a normalization factor after some probability amplitudes are set to zero due to measurement.
Unfortunately, a normalization factor may not be algebraically represented by Eq.~\eqref{eq:algebraic}.
Hence, we modify Eq.~\eqref{eq:algebraic} as
\begin{equation}\label{eq34}
s\times\alpha = s \times \frac{1}{{\sqrt{2}}^k}(a{\omega}^3+b{\omega}^2+c{\omega}+d),
\end{equation}
where $s\in\mathbb{R}$ is a normalization factor for measuring some qubit(s).
Note that the potential precision loss in this final simulation step of measurement is inevitable because we have to represent the answer probability using a floating point number anyway.
On the other hand, for measuring multiple qubits, having one measurement on all of the interested qubits yields less precision loss than a sequence of measurements on the qubits one at a time.
The former avoids the need of normalization (at the cost of exploring exponential number of outcomes), while the latter requires normalization several times.
If our interested query is about the probability of a particular outcome, e.g., $\Pr[q_0q_1q_2 = \ket{000}]$, rather than the probabilities of all possible outcomes, then the former is more preferred than the latter.

By replacing $\alpha$ with $s\times\alpha$ in Eq.~\eqref{eq4}, we derive
\begin{equation*}
\displaystyle\sum_{i \in \{0,1\}^n \mathrm{\ with\ } q=j} \abs{s\alpha_{i}}^{2} = s^2\displaystyle\sum_{i \in \{0,1\}^n \mathrm{\ with\ } q =j} \abs{\alpha_{i}}^{2},
\end{equation*}
where $j \in \{0, 1\}$.
Therefore, to compute the current probabilities after normalization, we can simply multiply $s^2$ with the accumulated probabilities of the current nodes.
Hence the accumulated probabilities need not be recomputed due to normalization.

Assume we would like to measure multiple qubits one at a time.
We make the BDD variables of the interested qubits on top and follow the measurement order in $F$.
Without loss of generality let the order be $q_0, \ldots, q_{n-1}$.
Then, the first qubit to be measured is $q_0$, which is the root node of $F$.
As shown in the left part of Fig.~\ref{fig_big_BDD}, $p_{00}$ and $p_{01}$ are the accumulated probabilities of the 0-child and 1-child of the root node, respectively.
Assume we obtain $q_0=\ket{1}$ after measurement, as shown in the right part of Fig.~\ref{fig_big_BDD}, the amplitudes with $q_0=\ket{0}$ can easily be set to zero by connecting the 0-edge of $q_0$ to the constant-0 node, and $s$ is updated from $1$ to $1/\sqrt{p_{01}}$ to satisfy the normalization constraint.
We repeat the same procedure to measure the other qubits following the order of variables.
In fact, when measuring $q_i$, there exists only one node labelled $q_i$.

It is worth noting that, when some qubits are to be measured, the order of measuring them is immaterial.
This freedom allows BDD reordering to be performed to reduce BDD size.
The only requirement is to make the qubit variables to be measured above the qubit variables not to be measured, and the encoding variables below all the qubit variables.

\section{Experimental Results} \label{sec:exp}
The proposed simulation algorithm was implemented in C++ in the \texttt{ABC} system \cite{ABC} and used \texttt{CUDD} \cite{CUDD} as the underlying BDD package.
For the dynamic variable reordering heuristic, the implementation of \cite{reorder} in \texttt{CUDD} was used.
In our implementation, the size of the integers $r$ was initially set to 32.
When overflows were detected, extra BDDs were allocated for each integer.
To reduce the precision loss due to probability calculation in measurement, we used the \texttt{GNU MPFR} library \cite{mpfr} to increase the precision of the floating point numbers.
In the experiments, the state-of-the-art simulator \texttt{DDSIM} (version v1.0.1a) \cite{zulehner2019advanced,QMDD2}, which is based on QMDD \cite{QMDD1}, was compared. All experiments were conducted on a server with Intel(R) Xeon(R) Silver 4210 CPU @ 2.20GHz, 30.7GB RAM. %and 20 cores (40 threads).
The time-out (TO) limit was set to 7200 seconds, and the memory-out (MO) limit was set to 2GB for each case.
The evaluation was performed on four sets of benchmarks, including 1) randomly generated benchmarks, 2) \texttt{RevLib} benchmarks  \cite{revlib} and its variants with entanglements, 3) quantum algorithm benchmarks for entanglement (Bell state preparation) and Bernstein–Vazirani (BV) algorithm \cite{Bernstein-Vazirani}, and 4)
supremacy benchmarks from Google \cite{supremacy18}.

%Since random circuits do not have structure that can be exploited by classical algorithms, they are a representative choice to evaluate the performances of simulators.
For the first set of benchmarks, we randomly generated circuits of various qubit sizes (40, 80, 120, 160, 200, 300, 400, 500) using all the supported gates, but excluding Rx($\pi/2$) and Ry($\pi/2$) as they exhibit similar effects as the H-gate.
The ratio of $\#gates:\#qubits$ was fixed to 3:1, and 10 circuits were generated for each size.
In building a circuit, we first inserted an H-gate to every qubit (so to impose state superposition in the beginning), and then inserted the targeted number of gates into the circuit by picking every gate uniformly at random from the mentioned gate set and applied it to some qubit(s) selected uniformly at random.

The results on the random circuits are shown in Table~\ref{tb:rand}, where Column~1 lists $\#qubits$, Column~2 lists $\#gates$, and Columns~3 and 4 list the average runtime and the numbers of TO/MO/error/segmentation fault cases of \texttt{DDSIM}, respectively, and Columns~5 and 6 list the similar information of our simulator.
The term `error' indicates a numerical error happened if all state probabilities do not sum to 1 due to precision loss.
The message `failed' means all 10 cases cannot be simulated successfully.
The reported runtime was only averaged over success cases.

From Table~\ref{tb:rand}, we see that \texttt{DDSIM} fails to simulate circuits with 120 or more qubits due to TO, MO, numerical errors, or segmentation faults.
In contrast, our method tends to be much more robust and scalable when simulating the considered random circuits.
\texttt{DDSIM} yielded 13 MO and 30 error cases.
On the contrary, our method produced no such cases.
In this benchmark setting, our simulator is superior to DDSIM in runtime, memory efficiency, and accuracy.

For the second benchmark set, we took reversible circuits \texttt{RevLib} \cite{revlib} for experiments.
However, because most of the circuits from \texttt{RevLib} are converted from classical circuits, they do not exhibit quantum effects and can be simulated efficiently.
(In the simulation we assumed random initial values for inputs whose initial values are not specified.)
To make the circuits more interesting with quantum effects, we modified the original circuits by inserting H-gates to the inputs whose initial values are not specified in the original circuit such that we create superposition states in the beginning.
The results on the \texttt{RevLib} benchmarks are shown in Table~\ref{tb:rev}, where Column~1 lists the circuit name, Column~2 lists the corresponding $\#qubits$, Columns~3-5 list the $\#gates$ before the modification and the results on the original circuits, and similarly, Columns~6-8 list the $\#gates$ after the modification and the results on the modified circuits.
As one can see from Table~\ref{tb:rev}, both \texttt{DDSIM} and our method can simulate the circuits of classical functionalities efficiently.
When the modified circuits are considered, \texttt{DDSIM} suffered mostly from MO when simulating the modified circuits.
For those circuits that \texttt{DDSIM} cannot simulate successfully, our method can simulate them within the timeout limit.\footnote{There are some \texttt{RevLib} benchmarks that our method cannot simulate successfully due to timeout. However, \texttt{DDSIM} fails on those circuits,
too.}
To further investigate the MO cases of \texttt{DDSIM}, we performed a case study on \texttt{callif\_32\_439} and removed the MO limit.
Nevertheless, \texttt{DDSIM} still cannot finish simulation within TO limit, and the memory usage grows to 9.72GB upon TO.

For the third benchmark set, we collected quantum algorithm circuits, including the entanglement and BV circuits.\footnote{There are other quantum algorithm circuits, such as QFT and Shor algorithms.
However, they involve unitary operators not algebraically representable and are excluded from our experiments.}
The results are shown in Table~\ref{tb:bv}, where Column~1 lists $\#qubits$, Columns~2-4 and 5-7 list the $\#gates$ and runtime information for entanglement and BV circuits, respectively.
For the entanglement circuits, \texttt{DDSIM} encountered MO at $\#qubits = 10000$, whereas our method finishes within 67 seconds.
For the BV circuits, \texttt{DDSIM} encountered numerical errors and segmentation faults for the $\#qubits \geq 90$ cases, whereas our method finishes in hundreds of seconds when simulating circuits with thousands of qubits.
We note that the entanglement circuits belong to the category of stabilizer circuits, which are known efficiently simulatable by classical computers \cite{Aaronson_2004}.
When \texttt{CHP} \cite{CHP}, a simulator based on \cite{Aaronson_2004} dedicated to stabilizer circuit simulation, is applied, the entanglement circuit with $\#qubits = 10000$ can be simulated in 6.7 seconds.
It is not surprising as \texttt{CHP} exploits additional circuit properties for fast simulation.
On the other hand, BV circuits are beyond the stabilizer circuit category and cannot be simulated by \texttt{CHP}.

For the fourth benchmark set, we took the random circuits proposed by Google for showing quantum supremacy \cite{supremacy18}.\footnote{The circuits were downloaded from \url{https://github.com/sboixo/GRCS}
under the directory "inst/rectangular/cz\_v2".}
These circuits are meant to create highly entangled states to make them challenging to simulate by classical computers.
As the circuits are too difficult to simulate, we simplified the circuits with depth 10 by reducing their depths to 5.
The results are shown in Table~\ref{tb:grcs}, where all columns list the same information as that in Table~\ref{tb:rand}, except that memory usage information is additionally reported and the numbers of error and segmentation fault cases are all 0 and omitted.
Similar to the experiments of Table~\ref{tb:rand}, we have 10 random circuits for each qubit size (in each row in Table~\ref{tb:grcs}), and the runtime was averaged over the cases that are simulated successfully.
On the other hand, the memory usage was averaged over all 10 cases including TO and MO cases.
The obtained results show that, \texttt{DDSIM} and our method suffer from MO and TO, respectively, when simulating $\#qubits \geq 42$.
As can be observed, \texttt{DDSIM} took smaller average runtime compared to ours.
However, by examining circuits with some specific $\#qubits$, we can observe that our method possibly simulates more cases than \texttt{DDSIM}.
For example, when $\#qubits = 49$, our method simulates 9 cases, whereas \texttt{DDSIM} simulates only 4 cases.
Overall, \texttt{DDSIM} simulates 74 out of the total 120 cases, whereas our method simulates 77 out of 120 cases. Furthermore, our method clearly outperforms \texttt{DDSIM} in terms of the memory usage.
%In contrast with over 1.5GB took by \texttt{DDSIM}, our method requires only up to 348MB.
This challenging benchmark set motivates us for further investigation and improvement.

\begin{table}[ht]
\caption{Results on Random Circuits.}
\resizebox{\columnwidth}{!}{
\begin{tabular}{|r|r|r|c|r|c|}
\hline
\multicolumn{1}{|c|}{\multirow{2}{*}{\#Qubits}} & \multicolumn{1}{c|}{\multirow{2}{*}{\#Gates}} & \multicolumn{2}{c|}{DDSIM}                     & \multicolumn{2}{c|}{Ours}                      \\ \cline{3-6}
\multicolumn{1}{|c|}{}                          & \multicolumn{1}{c|}{}                         & \multicolumn{1}{c|}{Time(s)} & TO/MO/err./seg. & \multicolumn{1}{c|}{Time(s)} & TO/MO/err./seg. \\ \hline
40                                              & 120                                           & 8.67                         & 0/0/0/0         & 0.82                         & 0/0/0/0         \\
80                                              & 240                                           & 502.29                       & 2/6/0/0         & 14.62                        & 0/0/0/0         \\
120                                             & 360                                           & failed                       & 3/5/2/0         & 473.26                       & 0/0/0/0         \\
160                                             & 480                                           & failed                       & 0/2/8/0         & 617.38                       & 2/0/0/0         \\
200                                             & 600                                           & failed                       & 0/0/10/0        & 297.48                       & 1/0/0/0         \\
300                                             & 900                                           & failed                       & 0/0/10/0        & 647.98                       & 5/0/0/0         \\
400                                             & 1200                                          & failed                       & 0/0/0/10        & 2532.65                      & 5/0/0/0         \\
500                                             & 1500                                          & failed                       & 0/0/0/10        & 2485.64                      & 9/0/0/0         \\ \hline
\end{tabular}}
\label{tb:rand}
\end{table}

\begin{table}[ht]
\caption{Results on Revlib Circuits.}
\resizebox{\columnwidth}{!}{
\begin{tabular}{|l|r|r|r|r|r|r|r|}
\hline
\multicolumn{1}{|c|}{}                            & \multicolumn{1}{c|}{}                           & \multicolumn{3}{c|}{Original}                                                                           & \multicolumn{3}{c|}{Modified}                                                                           \\ \cline{3-8}
\multicolumn{1}{|c|}{}                            & \multicolumn{1}{c|}{}                           & \multicolumn{1}{c|}{}                          & \multicolumn{2}{c|}{Time(s)}                           & \multicolumn{1}{c|}{}                          & \multicolumn{2}{c|}{Time(s)}                           \\ \cline{4-5} \cline{7-8}
\multicolumn{1}{|c|}{\multirow{-3}{*}{Benchmark}} & \multicolumn{1}{c|}{\multirow{-3}{*}{\#Qubits}} & \multicolumn{1}{c|}{\multirow{-2}{*}{\#Gates}} & \multicolumn{1}{c|}{DDSIM} & \multicolumn{1}{c|}{Ours} & \multicolumn{1}{c|}{\multirow{-2}{*}{\#Gates}} & \multicolumn{1}{c|}{DDSIM} & \multicolumn{1}{c|}{Ours} \\ \hline
\_443                                             & 261                                             & 1251                                           & 0.31                       & 0.17                      & 1317                                           & MO                         & 166.55                    \\
add64\_184                                        & 193                                             & 256                                            & 0.05                       & 0.05                      & 385                                            & 0.14                       & 0.09                      \\
apex2\_289                                        & 498                                             & 1746                                           & 0.59                       & 0.46                      & 1785                                           & {\color[HTML]{333333} MO}  & 4610.48                   \\
callif\_32\_439                                   & 130                                             & 561                                            & 0.08                       & 0.07                      & 626                                            & MO                         & 3.68                      \\
cps\_292                                          & 923                                             & 2763                                           & 3.28                       & 1.10                       & 2787                                           & {\color[HTML]{333333} MO}  & 5059.14                   \\
cpu\_alu\_16bit\_400                              & 405                                             & 6487                                           & 0.83                       & 0.28                      & 6526                                           & MO                         & 718.48                    \\
cpu\_control\_unit\_402                           & 392                                             & 1351                                           & 0.24                       & 0.09                      & 1514                                           & MO                         & 1569.36                   \\
cpu\_register\_32\_405                            & 328                                             & 597                                            & 0.35                       & 0.07                      & 890                                            & 0.53                       & 0.35                      \\
e64-bdd\_295                                      & 195                                             & 387                                            & 0.1                        & 0.06                      & 452                                            & 3.08                       & 3.03                      \\
ex5p\_296                                         & 206                                             & 647                                            & 0.14                       & 0.08                      & 655                                            & 1.43                       & 11.54                     \\
hwb9\_304                                         & 170                                             & 699                                            & 0.13                       & 0.07                      & 708                                            & 5.84                       & 12.53                     \\
lu\_326                                           & 299                                             & 571                                            & 0.19                       & 0.25                      & 637                                            & MO                         & 6.20                       \\
nestedif2\_32\_445                                & 263                                             & 854                                            & 0.24                       & 0.23                      & 920                                            & MO                         & 291.26                    \\
pdc\_307                                          & 619                                             & 2080                                           & 1.28                       & 0.68                      & 2096                                           & {\color[HTML]{333333} MO}  & 5856.65                   \\
spla\_315                                         & 489                                             & 1709                                           & 0.72                       & 0.35                      & 1725                                           & {\color[HTML]{333333} MO}  & 1925.49                   \\
varops\_32\_447                                   & 224                                             & 1305                                           & 0.29                       & 0.12                      & 1401                                           & MO                         & 3271.49                   \\ \hline
\end{tabular}}
\label{tb:rev}
\end{table}

\iffalse
\subsection{RevLib as Benchmarks}
To evaluate the performance using real circuits, we took circuits from a well-established online resource \texttt{RevLib} \cite{revlib}. However, most of original circuits from \texttt{RevLib} can be simulated in trivial running time since they result in rather small sizes of decision diagrams. To this end, we modified the original circuits by applying Hadamard gates in the beginning on the qubits whose initial states are not specified. This means that inputs of the modified circuits were the superpositions of all possible input basis states.

The results on these benchmarks are shown in table X, where column ... Note that, we only present circuits 1) whose modified versions were simulated successfully by at least one simulator, 2) with more than a hundred qubits (since all circuits with less than a hundred qubits can be simulated within ten seconds), and 3) with the biggest number of qubits in their family (e.g., since "cpu\_register\_16\_452" has 167 qubits and "cpu\_register\_32\_333" has 327 qubits, only the second one is presented). No QMDD finishes but BDD doesn't finish ...
\fi

% \usepackage{multirow}
\begin{table}[ht]
\caption{Results on Quantum Algorithm Circuits.}
\begin{tabular}{|r|r|r|r|r|r|r|}
\hline
\multicolumn{1}{|c|}{\multirow{3}{*}{\#Qubits}} & \multicolumn{3}{c|}{Entanglement}                                                                      & \multicolumn{3}{c|}{BV}                                                                                \\ \cline{2-7}
\multicolumn{1}{|c|}{}                          & \multicolumn{1}{c|}{\multirow{2}{*}{\#Gates}} & \multicolumn{2}{c|}{Time(s)}                           & \multicolumn{1}{c|}{\multirow{2}{*}{\#Gates}} & \multicolumn{2}{c|}{Time(s)}                           \\ \cline{3-4} \cline{6-7}
\multicolumn{1}{|c|}{}                          & \multicolumn{1}{c|}{}                         & \multicolumn{1}{c|}{DDSIM} & \multicolumn{1}{c|}{Ours} & \multicolumn{1}{c|}{}                         & \multicolumn{1}{c|}{DDSIM} & \multicolumn{1}{c|}{Ours} \\ \hline
80                                              & 80                                            & 0.01                       & 0.02                      & 239                                           & 0.02                       & 0.06                      \\
90                                              & 90                                            & 0.01                       & 0.01                      & 269                                           & error                      & 0.07                      \\
100                                             & 100                                           & 0.01                       & 0.01                      & 299                                           & error                      & 0.07                      \\
500                                             & 500                                           & 0.19                       & 0.19                      & 1499                                          & seg. fault                 & 4.48                      \\
1000                                            & 1000                                          & 0.87                       & 0.99                      & 2999                                          & seg. fault                 & 116.80                     \\
5000                                            & 5000                                          & 46.21                      & 15.2                      & 14999                                         & seg. fault                 & 186.93                    \\
10000                                           & 10000                                         & MO                         & 66.95                     & 29999                                         & seg. fault                 & 595.74                    \\ \hline
\end{tabular}
\label{tb:bv}
\end{table}

% \usepackage{multirow}
\begin{table}[ht]
\caption{Results on Google Supremacy Circuits.}
\resizebox{\columnwidth}{!}{
\begin{tabular}{|r|r|r|r|c|r|r|c|}
\hline
\multicolumn{1}{|c|}{\multirow{2}{*}{\#Qubits}} & \multicolumn{1}{c|}{\multirow{2}{*}{\#Gates}} & \multicolumn{3}{c|}{DDSIM}                                          & \multicolumn{3}{c|}{Ours}                                           \\ \cline{3-8}
\multicolumn{1}{|c|}{}                          & \multicolumn{1}{c|}{}                         & \multicolumn{1}{c|}{Time(s)} & \multicolumn{1}{c|}{Mem(MB)} & TO/MO & \multicolumn{1}{c|}{Time(s)} & \multicolumn{1}{c|}{Mem(MB)} & TO/MO \\ \hline
16                                              & 61                                            & \textless{}0.01              & 86.99                        & 0/0   & 0.14                         & 62.91                        & 0/0   \\
20                                              & 79                                            & 0.01                         & 86.99                        & 0/0   & 1.09                         & 76.13                        & 0/0   \\
25                                              & 99                                            & 0.01                         & 87.00                        & 0/0   & 4.38                         & 98.66                        & 0/0   \\
30                                              & 120                                           & 0.01                         & 87.00                        & 0/0   & 4.19                         & 100.30                       & 0/0   \\
36                                              & 144                                           & 0.58                         & 90.87                        & 0/0   & 25.86                        & 111.12                       & 0/0   \\
42                                              & 169                                           & 0.16                         & 871.63                       & 0/4   & 938.06                       & 136.50                       & 1/0   \\
49                                              & 202                                           & 4.50                         & 1274.99                      & 0/6   & 2095.56                      & 248.86                       & 1/0   \\
56                                              & 225                                           & 93.24                        & 944.89                       & 0/3   & 1478.19                      & 249.74                       & 5/0   \\
64                                              & 254                                           & 10.66                        & 1286.31                      & 0/6   & 3339.24                      & 335.49                       & 6/0   \\
72                                              & 290                                           & 613.29                       & 1791.28                      & 0/8   & failed                       & 373.52                       & 10/0  \\
81                                              & 328                                           & 2.22                         & 1795.43                      & 1/8   & failed                       & 278.66                       & 10/0  \\
90                                              & 365                                           & failed                       & 1745.23                      & 3/7   & failed                       & 347.84                       & 10/0  \\ \hline
\end{tabular}
\label{tb:grcs}}
\end{table}

\section{Conclusions} \label{sec:concl}
In this paper, we have presented a new quantum circuit simulator that outperforms the state-of-the-art in both scalability and accuracy.
The algebraic representation, bit-slicing technique, and quantum gate formula pre-characterization together enable the success.
For future work, we would like to investigate the supremacy benchmarks further and identify points for improvements.

\section*{Acknowledgment}
This work was supported in part by the Ministry of Science and Technology of Taiwan under grant MOST 108-2218-E-002-073.
The authors are grateful to Stefan Hillmich for answering questions about the DDSIM package.

\bibliographystyle{IEEEtran}
\bibliography{reference}

\end{document}